\documentclass{emulateapj}

\newcommand{\etal}{{\em et al.}}

\newcommand\fsg{$f_{\rm SG,H}~$}
\newcommand\fsge{f_{\rm SG,H}}
\newcommand\fsgcs{$F_{\rm SG,GCS}^{now}~$}
\newcommand\fsgcse{F_{\rm SG,GCS}^{now}}
\newcommand\phihalo{\Phi}
\def\simgt{\lower.5ex\hbox{$\; \buildrel > \over \sim \;$}}
\def\simlt{\lower.5ex\hbox{$\; \buildrel < \over \sim \;$}}

\shorttitle{Globular cluster second-generation stars in
  the Galactic halo}
\shortauthors{Vesperini et al.}

\begin{document}

\title{The fraction of globular cluster second-generation stars in
  the Galactic halo}

\author{Enrico Vesperini, Stephen L.W. McMillan}
\affil{Department of Physics, Drexel University, Philadelphia, PA 19104}
\author{Francesca D'Antona}
\affil{INAF, Osservatorio Astronomico di Roma, via di Frascati 33,
  I-00040 Monteporzio, Italy}
\author{Annibale D'Ercole}
\affil{INAF, Osservatorio Astronomico di Bologna, via Ranzani 1,
  I-40127 Bologna, Italy}
\submitted{Accepted for publication in The Astrophysical Journal, Letters}
\begin{abstract}
  Many observational studies have revealed
  the presence of multiple stellar generations in Galactic
  globular clusters.  These studies suggest that second-generation
  stars make up a significant fraction of the current mass of globular
  clusters, with the second-generation mass fraction ranging from
  $\sim$50 to 80\% in individual clusters.  In this Letter we carry out
  hydrodynamical simulations to explore the dependence of the mass of
  second-generation stars on the initial mass and structural
  parameters and stellar initial mass function (IMF) of the parent
  cluster.  We then use the results of these 
  simulations to estimate the fraction $\fsge$ of the mass of the
  Galactic stellar halo composed of second-generation stars that
  originated in globular clusters.  We study the dependence of \fsg on
  the parameters of the initial mass function of the Galactic globular
  cluster system. For a broad range of initial conditions, we find
  that the fraction of mass of the Galactic stellar halo in
  second-generation stars is always small, $\fsge<4-6\%$ for a
  Kroupa-1993 IMF and  $\fsge<7-9\%$ for a Kroupa-2001 IMF.
\end{abstract}

\section{Introduction} 
\label{sec:introduction}
During the last decade, spectroscopic and photometric studies have
provided strong evidence for multiple stellar populations in a number
of Galactic globular clusters. In contrast to the standard formation
scenario, according to which globular clusters are 'simple stellar
populations' composed of stars all with the same age and chemical
composition, these studies have revealed star-to-star variations in
the abundances of light elements in main sequence and red giant stars,
suggesting a second stellar generation (hereafter SG) that must have
formed out of gas processed through a first generation (hereafter FG)
of stars. SG stars are identified by their anomalous abundances of Na,
O, Mg and Al relative to normal halo field stars of similar
metallicity (see e.g. Gratton et al. 2004, Carretta et al. 2009a,b).

Accurate photometric studies of a few globular clusters have
strengthened the evidence for multiple populations by showing their
fingerprints (multiple main sequences and sub-giant branches) in the
color-magnitude diagram (see e.g. Lee et al. 1999, Pancino et
al. 2000, Bedin et al. 2004, Piotto et al. 2007, Milone et al. 2008,
Marino et al. 2008, Anderson et al. 2009).  The most striking main
sequence anomalies are populations with helium abundances much larger
than the standard Big-Bang value expected to be universally present in
the oldest stars.  The helium overabundance is interpreted as an
evolutionary by-product of the progenitors of the anomalous SG stars.
The leading models for the source gas from which SG stars formed involve
rapidly rotating massive stars (e.g. Decressin et al. 2007; see also
de Mink et al. 2009 for a more recent work suggesting massive binaries
as a possible polluting source), or 
intermediate-mass asymptotic giant branch (AGB) stars (e.g. Ventura et
al. 2001; see also Renzini 2008 for a review and references therein).

Prior to the discovery of the multiple main sequences, the presence of
stars with helium anomalies had been predicted based on the very
extended horizontal branches of some clusters (D'Antona et al. 2002,
D'Antona and Caloi 2004).  Analysis of horizontal branch morphologies
of several clusters also led to the first suggestion that the second
stellar generation could account for a high fraction---from 30\% to
100\%---of the total number of stars (D'Antona \& Caloi 2008).  While
the high-quality photometric data necessary to detect multiple
populations have been so far obtained for only a few clusters, a
recent spectroscopic survey of about 2000 stars in 19 Galactic
globular clusters has shown evidence of multiple populations in all
the clusters observed, and in all cases SG stars account for a
significant fraction ($50-80\%$) of the cluster mass (Carretta et
al. 2009a).

D'Ercole {\etal} (2008) studied the formation and dynamical evolution
of multiple populations in globular clusters, and found that two
distinct evolutionary processes can cause stars to escape and populate
the Galactic stellar halo during cluster dynamical evolution.  At
early times, stars escape over the tidal boundary as the cluster
expands due to the loss of primordial gas and FG SN ejecta.  Since SG
stars form preferentially in the central regions of the cluster,
almost all of the escapers at this stage are FG stars.  The result is
that the SG fraction among bound cluster stars increases dramatically.
As the cluster evolution continues into the phase dominated by
two-body relaxation, the two populations mix and the fractional escape
rates of the FG and SG stars due to evaporation tend to equalize,
stabilizing the SG fraction (see also Decressin et al. 2008 for another
study of the mixing process).  Thus
the vast majority of escapers from 
the cluster are FG stars.  Two recent spectroscopic studies (Carretta
et al. 2010, Martell \& Grebel 2010) have found that the vast majority
of halo stars studied have abundances typical of FG stars in clusters;
only about 1.5--2.5\% of the stars are Na-rich (Carretta et al. 2010) and
CN-strong (Martell \& Grebel 2010), and hence classifiable as SG
stars.

Many general studies have addressed the evolution of the Galactic
globular cluster system and the dispersal of cluster stars in the halo
(see e.g. Kroupa \& Boily 2002, Baumgardt et al. 2008, Portegies Zwart
et al. 2010, Vesperini 2010 and references therein).  However, this
Letter is specifically focused on SG  
stars.  Our goals here are to estimate both the number of SG stars
that may have formed in the Galactic globular cluster system and the
fraction, $\fsge$, of SG stars formed in globular clusters that now populate
the Galactic stellar halo, and to illustrate the link between SG
formation, globular cluster evolution and the formation history of the
Galactic halo. In \S\ref{sec:hydro}, we present the results of
hydrodynamical simulations exploring the dependence of the mass of SG
stars formed on cluster initial properties. In \S\ref{sec:sghalo} we
introduce our numerical framework to estimate the fraction of SG stars
in the halo, and discuss its theoretical and observational
ingredients. In \S\ref{sec:res} we present and discuss our results.

\section{Mass of second-generation stars as a function of a cluster
  initial properties}
\label{sec:hydro}
In order to calculate the total mass of SG stars formed in a cluster
as a function of cluster initial properties, we have carried out a
series of one-dimensional hydrodynamical simulations modeling the SG
formation process. The technical details of our simulations are
described in D'Ercole {\etal} (2008). We outline here only the aspects
relevant to this Letter.

We assume that the FG cluster initially follows a King (1962) density
profile
\begin{equation}
    \rho=\rho_0\left[1+\left({r\over r_c}\right)^2\right]^{-1.5}
\end{equation} 
out to some truncation radius $R$. Stellar masses are distributed in the range
$0.1<m/m_{\odot}<100$  according to either a Kroupa et
al. (1993; hereafter K93) or a Kroupa (2001; hereafter K01) stellar IMF.
We model the formation of
SG stars from the ejecta of AGB stars and assume that the SG formation
phase extends from about 30 Myr after the FG cluster forms until $\sim100$
Myr (should the SG star formation epoch extend to $\sim 150$ Myr, and hence
include ejecta of stars down to 4 $M_{\odot}$, the SG mass can be
obtained by multiplying the values presented below by a factor $\sim
1.4$).  As discussed in D'Ercole {\etal} (2008) and D'Ercole {\etal}
(2010) in order to explain the chemical properties of
SG stars it is necessary to invoke the presence of pristine gas mixed
in with the matter from which SG stars formed (see also
Pflamm-Altenburg \& Kroupa 2009 for a study of the possible accretion
of gas from the ISM during the SG formation process). As shown by
D'Ercole et al. (2008, 2010), a mass of pristine gas equal  
to about 50\% of the total mass of SG star-forming gas is required to
reproduce the abundance patterns (Na-O, Mg-Al anticorrelations and the
helium distribution function) observed in the massive cluster NGC2808
and in the low-mass cluster M4 (see also Ventura {\etal} 2009, Ventura
\& D'Antona 2009).  

However, additional systematic studies of the
chemical properties of the individual clusters for which spectroscopic
data are available will be required to explore the allowed mass range
of pristine gas and any possible dependences on protocluster
conditions.  Here we simply assume that pristine material makes up
50\% of the total mass of gas forming the SG stars.  Results for
different pristine gas fractions can be easily obtained from those
presented in this Letter.
\begin{figure}    
\centering{
\includegraphics[width=8cm]{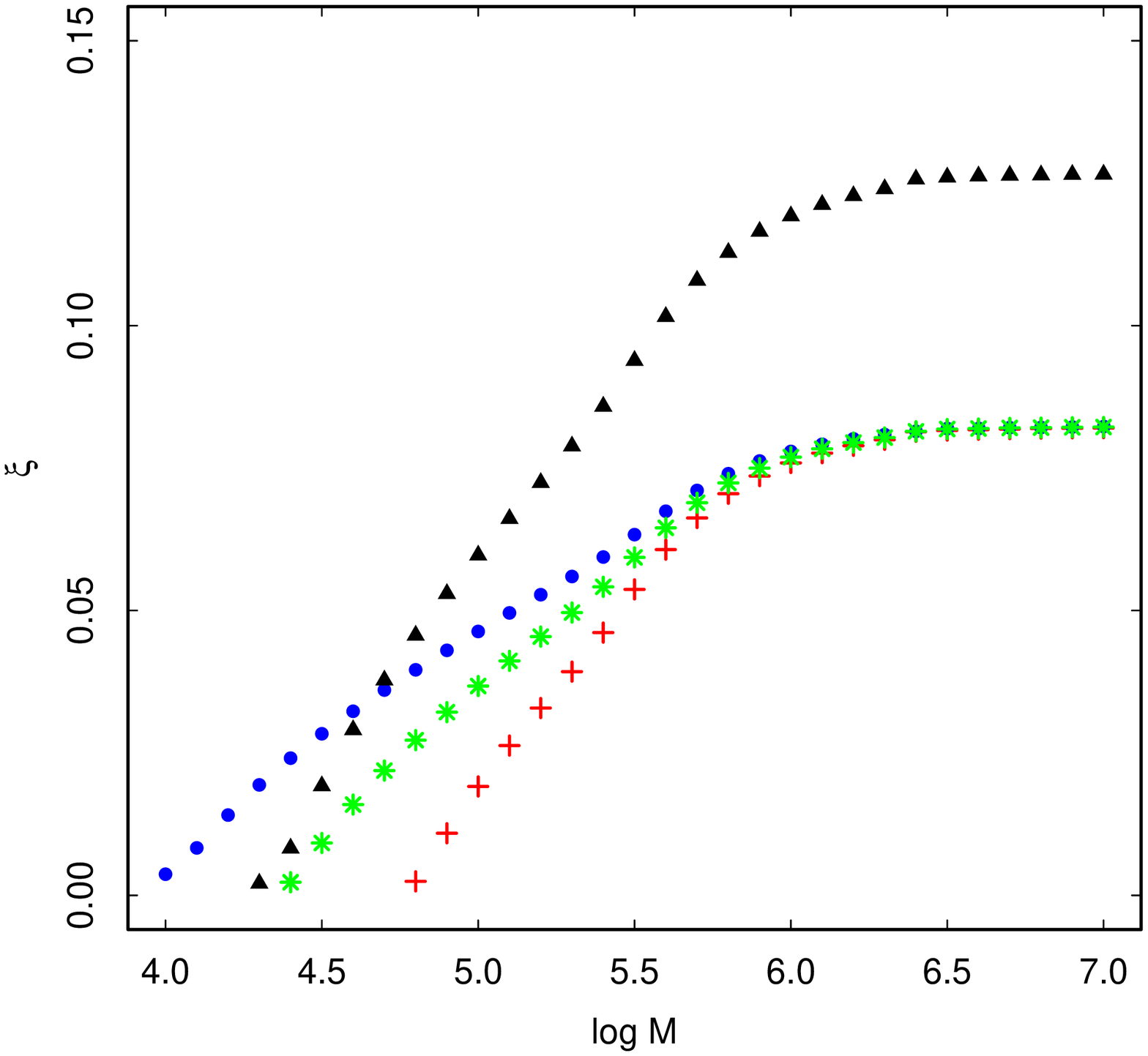}
\includegraphics[width=8cm]{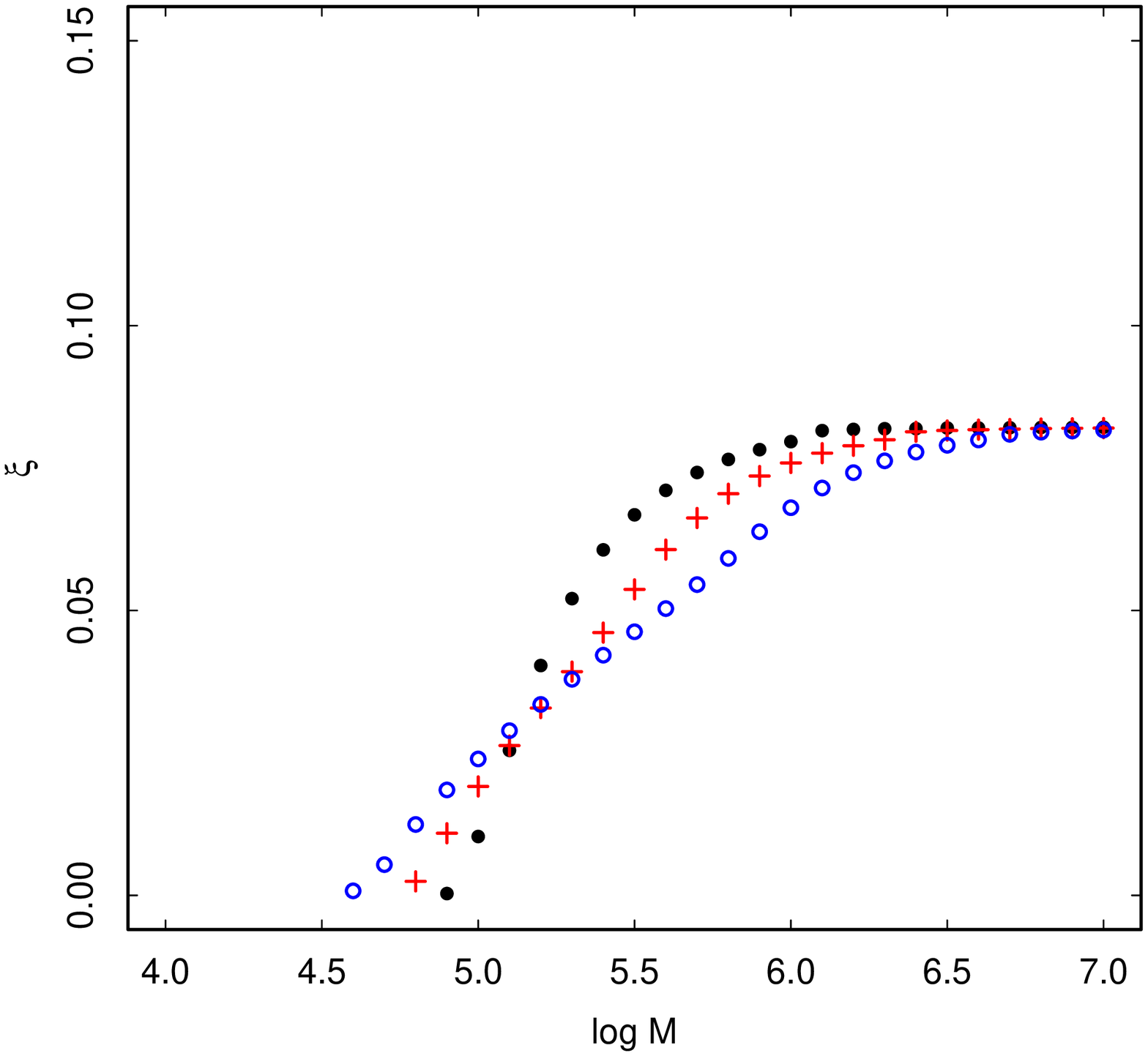}
}
\caption{Upper panel: Ratio $\xi$ of the total initial mass in SG stars to the
  total initial mass of the cluster $M$, as a function of $M$, for
  models with $R=40$ pc and $R_h=1$ pc, K93 IMF (filled circles), $R_h=2$ pc, K93 IMF (asterisks), $R_h=4$ pc, K93 IMF (crosses), $R_h=2$ pc, K01 IMF (filled triangles). 
Lower  panel: $\xi$ as a function of $M$, for models with $R_h=4$ pc, K93 IMF
 and different values of the cluster truncation radius $R$: $R=20$ pc (filled circles), $R=40$ pc (crosses),
  $R=80$ pc ( open circles). }
\label{fig:hydrofig} 
\end{figure}

The two panels of Fig.~\ref{fig:hydrofig} show the fractional mass of
SG stars formed, 
$\xi\equiv (M_{SG}/M)_{init}$, as a function of the total
cluster mass ($M$), for models with different combinations of values
of the half-mass 
radius ($R_h=1,~2,~4$ pc), of the   
truncation radius ($R=20$, 40, and 80 pc; these values encompass the
range of R for the majority   
of Galactic globular clusters, for which the modal value is $R\sim 40
$ pc; see e.g. Mackey \& van den Bergh 2005) and for a K93 or K01 IMF.
As the cluster mass 
 increases for given values of $R$ and $R_h$, so too do the cluster
escape speed, the fraction of AGB ejecta retained, and the total mass
of SG stars formed. For the structural parameters explored, only
clusters initially more massive than $\sim 10^{4.8}-10^5~M_{\odot}$ can
retain enough AGB ejecta to form an appreciable fraction of SG stars.
The difference between the values of $\xi$ obtained for the K93 and the
K01 IMF is a consequence of the fact that the K01 IMF contains a larger
fraction of the cluster mass in the mass range of the  polluting AGB stars.

\section{Second-generation stars in the Galactic stellar halo} 
\label{sec:sghalo}
\begin{figure}    
\centering{
\includegraphics[width=8cm]{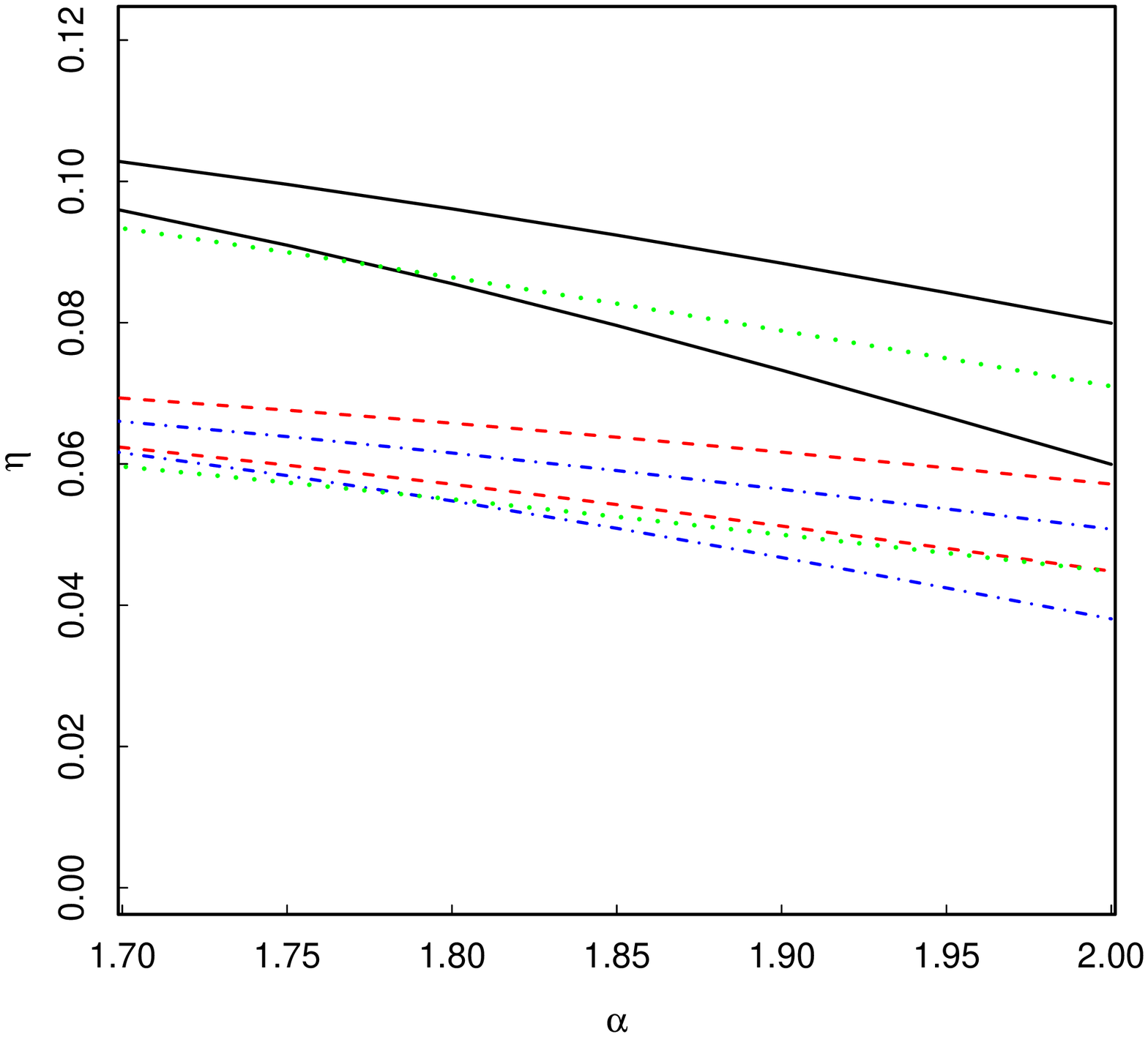}
\includegraphics[width=8cm]{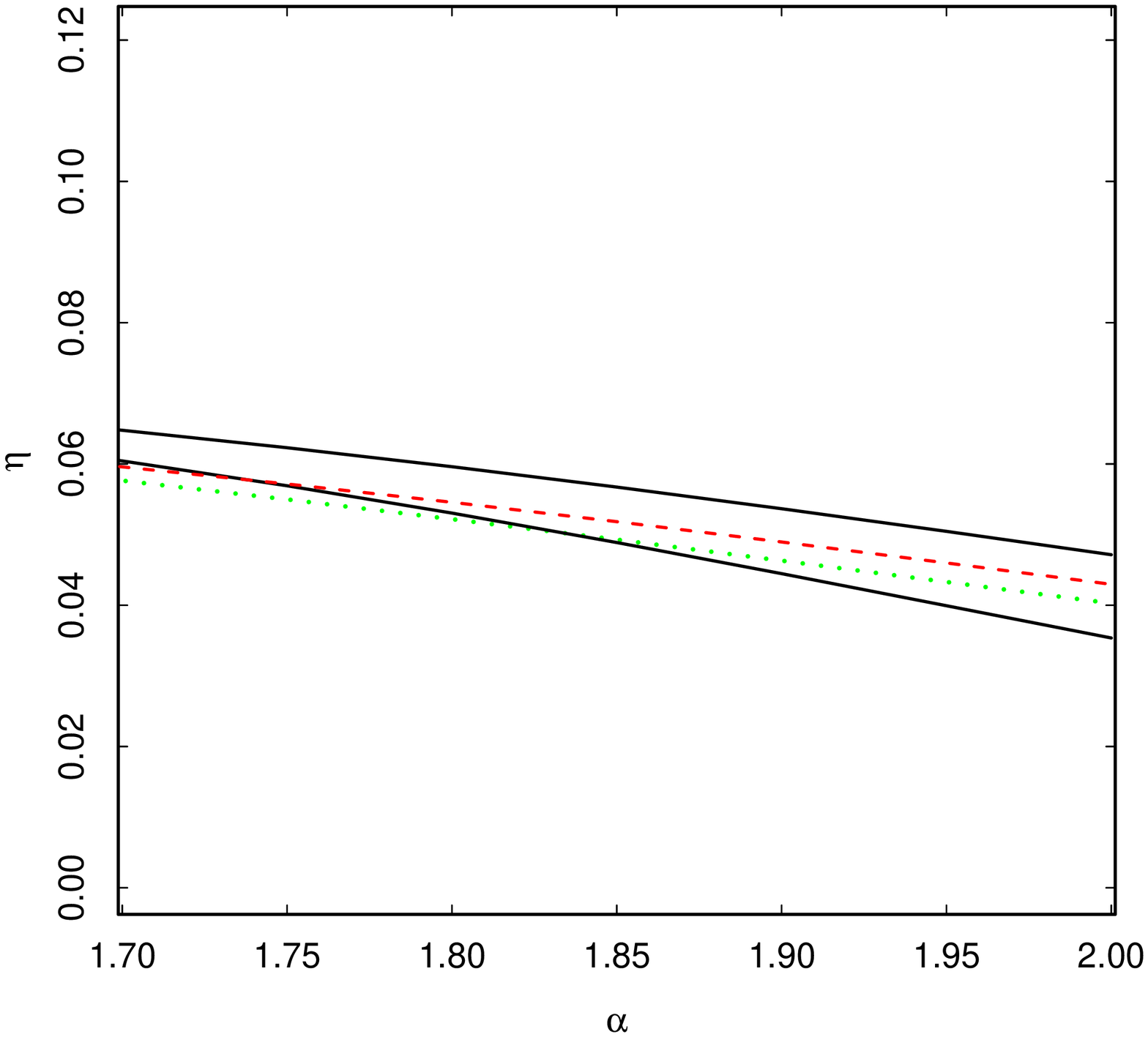}
}
\caption{Ratio, $\eta$, of the total initial mass of
SG stars in the Galactic globular cluster system to the total initial
mass of the cluster system as a function of the index
  $\alpha$ of the IGCMF. Different lines correspond to models with
  the following set of properties (IMF, $R_h$, $R$, $\log(M_C)$ (or  PL=power-law), $\log(M_{low})$): (upper panel)
(K01, 2, 40, PL, 4) upper black solid line;
(K01, 2, 40, PL, 3) lower black solid line;
(K01, 2, 40, 6.75, 4) upper green dotted line;
(K93, 1, 40, PL, 4) upper red dashed line;
(K93, 4, 40, PL, 4) lower red dashed line;
(K93, 2, 40, PL, 4) upper blue dot-dashed line;
(K93, 2, 40, PL, 3) lower blue dot-dashed line;
(K93, 2, 40, 6.75, 4) lower green dotted line. 
(Lower panel)
(K93, 4, 20, PL, 4) upper black solid line; 
(K93, 4, 20, PL, 3) lower black solid line; 
(K93, 4, 20, 6.75, 4) green dotted line;
(K93, 4, 80, PL, 4) red dashed line. 
  }
\label{fig:fsg1} 
\end{figure}
A complete model for the determination of the fraction of SG stars in
the Galactic halo would require accurate knowledge of the initial
global properties of the Galactic globular cluster system as well as
its dynamical history, the initial internal structure and subsequent
dynamical evolution of clusters with multiple stellar populations, and
a comprehensive theory of how the stellar halo formed.  Much work
remains to be done on all of these problems.  Considering the
uncertainties involved in all of these key ingredients, we elect to
proceed here in a simplified way, and our estimation of \fsg proceeds
as follows.

For a given value of the total mass of the Galactic globular
cluster system, $M_{GCS}^{init}$, and a given initial globular cluster system
mass function (hereafter IGCMF), we calculate the total mass of SG
stars formed in the cluster system, $M_{SG,GCS}^{init}$, using the
results of the hydrodynamical simulations described in the previous
section.  $M_{SG,GCS}^{init}$ may be written, in general, as an integral over
cluster structural properties
\begin{equation}
    M_{SG,GCS}^{init}=\int\int\int \xi(M,R_h,R)\, M f(M, R_h, R) \,dM dR_h dR,
\label{eq:msg1}
\end{equation}
where $f(M, R_h, R)$ is the joint distribution function of initial
cluster mass and structural parameters.  

We assume either a Schechter IGCMF (see e.g. Burkert \& Smith 2000)
\begin{equation}
    {dN\over dM} \propto M^{-\alpha}e^{-M/M_C} \hbox{~for
      $M_{low}<M<M_{up}$},
\label{eq:msg0}
\end{equation}
or a simple power-law (e.g. Whitmore 2003) 
\begin{equation}
    {dN\over dM} \propto M^{-\alpha} \hbox{~for
      $M_{low}<M<M_{up}$}.
\label{eq:msg2}
\end{equation}
With these assumptions, and for given values of $R_h$ and $R$ at the
time of SG formation ($30 \simlt t \simlt 100$ Myr),
Eq.\,\ref{eq:msg1} becomes 
\begin{equation}
    M_{SG,GCS}^{init} = \int_{M_{low}}^{M_{up}} \xi(M,R_h,R)\, M
			{dN\over dM} \,dM.
\label{eq:msg}
\end{equation}
Neglecting mass loss due to stellar evolution, the total mass of the
stellar halo in the form of SG stars, either from dissolved globular
clusters or evaporated from those remaining, is the difference between
$M_{SG,GCS}^{init}$ and the current total mass of SG stars still in
clusters, $M_{SG,GCS}^{now}$.

A systematic observational study of SG stars in Galactic globular
clusters is still lacking, and will require significant effort.
Currently the only study affording an estimate of the fraction of SG
stars in a significant number of Galactic clusters, as well as
exploration of possible trends of the SG fraction with other cluster
properties, is the spectroscopic survey of $\sim2000$ stars in 19
Galactic globular clusters carried out by Carretta {\etal} (2009a,b).
They find that all clusters studied host SG stars, and that the
fraction of individual cluster masses in SG stars ranges from about 50\%
to more than 80\%.  Hence we define the current cluster SG mass
fraction $\fsgcse$ by
\begin{equation}
    M_{SG,GCS}^{now} = \fsgcse M_{GCS}^{now},
\label{eq:msgnow}
\end{equation}
 where $M_{GCS}^{now}\sim 2\times
10^{7} M_{\odot}$ is the current total mass of halo Galactic globular
clusters (assuming $M/L=2$; Mackey \& van den Bergh 2005).

Most of the clusters studied are more massive than $\sim 10^5
M_{\odot}$, and it is not known whether such a large SG fraction is
common to all Galactic globular clusters or only to those currently
more massive than some threshold.  For purposes of this Letter we simply
assume as a reference value $\fsgcse \simeq 0.5$. Results for
different values of $\fsgcse$ 
can be very easily derived from those shown in this Letter.

To complete our calculation we need an estimate of $M_{GCS}^{init}$,
the initial total mass of all stars in globular clusters.  This is not
easily determined from observations of the current Galactic globular
cluster system.  Here again, we simply parametrize the normalization
of Eq.\,\ref{eq:msg0} or \ref{eq:msg2} by defining the parameter
$\phihalo=M_{GCS}^{init}/M_{halo}$ where we take $M_{halo}=10^9
M_{\odot}$ (see e.g. Freeman \& Bland-Hawthorn 2002).

The current fraction of SG stars in the stellar halo, $\fsge$, then is
\begin{equation}
    \fsge = {{M_{SG,GCS}^{init}-M_{SG,GCS}^{now}} \over M_{halo}}.
\label{eq:fsg}
\end{equation}
Finally, if we define $\eta$ as the ratio of the total initial mass of
SG stars in the Galactic globular cluster system to the total initial
mass of the cluster system
\begin{equation}
    \eta \equiv M_{SG,GCS}^{init}/M_{GCS}^{init},
\label{eq:eta}
\end{equation}
and use the definition of $\fsgcse$ in Eq.\,\ref{eq:msgnow} along with
the values of $M_{halo}$ and $M_{GCS}^{now}$ adopted above, we can
rewrite Eq.\,\ref{eq:fsg} as
\begin{equation}
    \fsge = \eta\phihalo - 0.02\fsgcse.
\label{eq:fsgb}
\end{equation}

The form of Eq.\,\ref{eq:fsgb} allows to easily identify the different
theoretical and observational ingredients needed for the determination
of $\fsge$.  We summarize them here.
\begin{itemize}
\item $\eta$, the ratio of the total initial mass of SG stars to the
  total initial mass of the Galactic globular cluster system,
  incorporates the dependency on the SG formation model (through the
  function $\xi(M,R_h,R)$), as well as on the stellar IMF, the cluster mass and
  structural parameter distribution functions (see Eq.\,\ref{eq:msg}
  above).
\item $\phihalo$, the ratio of the total initial mass of the Galactic
  globular cluster system to the current halo mass, is not known.  The
  simple linear scaling of \fsg with this parameter allows to easily
  explore the dependence of \fsg on the total initial mass of the
  Galactic globular cluster system.
\item $\fsgcse$, the fraction of the total current mass of the halo
  cluster population composed of SG stars, is a parameter to be
  determined observationally.  In our study, based on the
  spectroscopic survey of Carretta \etal (2009a,b), we adopt as a
  reference value $\fsgcse=0.5$.  However, in this case too, the
  simple linear scaling of \fsg with \fsgcs allows to easily calculate
  \fsg for other values of \fsgcs.
\end{itemize}

\section{Results and discussion}
\label{sec:res}

In Fig.\,\ref{fig:fsg1} we show the dependence of $\eta$ on the
power-law index $\alpha$ of the IGCMF. 
Different lines correspond to different pairs of values of $R_h$ and
$R$, a K93 or a K01 IMF and different values for the parameters of the
IGCMF ($M_C$ and $M_{low}$; $M_{up}$ is fixed at $10^7~M_{\odot}$).

Comparison of the results for models with different values of $R$ shows no
significant differences.  

An increase in $M_C$ and/or $M_{low}$ increases the fraction of the
total initial mass in clusters massive enough to form SG stars
($M\simgt 10^{4.8}-10^{5}$; see Fig. \ref{fig:hydrofig}), and hence
increases $\eta$.  Similarly, as $\alpha$ decreases the fraction of
total mass in massive clusters increases, and so does $\eta$.

For a fixed value
of $R$, decreasing the value of $R_h$ adopted has the obvious
consequence of increasing the fraction of AGB ejecta retained and the
amount of SG stars formed in lower-mass clusters. The absolute upper
limit ($\eta_{max} \simeq 0.08$ for a K93 IMF and $\eta_{max} \simeq 0.125$  for a
K01 IMF) would be attained by assuming
(unrealistically) that all clusters (including the low-mass ones)
were compact enough at the time of 
SG formation ($30\simlt t \simlt 100$ Myr) to retain all the AGB
ejecta and  reach a fraction of SG stars formed $\xi\simeq 0.08$ or
$\xi\simeq 0.125$ for a K93 and a K01 IMF respectively.
\begin{figure}
\centering{
\includegraphics[width=8cm]{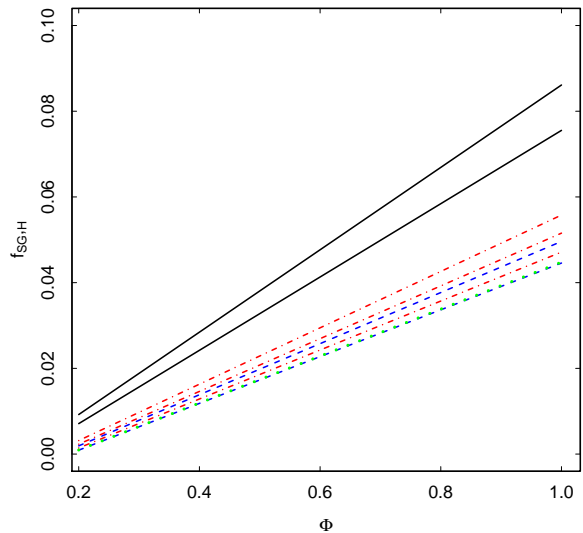}
}
\caption{\fsg versus $\phihalo$. A power-law IGCMF is assumed with
  $\alpha=1.8$. Different lines correspond to the following set of properties
(IMF, $R_h$, $R$, $\log(M_{low})$): 
(K01, 2, 40, 4) upper black solid line;
(K01, 2, 40, 3) lower black solid line;
(K93, 4, 20, 4) upper blue dashed line;
(K93, 4, 80, 4) lower blue dashed line;
(K93, 1, 40, 4) upper red dot-dashed line;
(K93, 2, 40, 4) middle red dot-dashed line;
(K93, 4, 40, 4) lower red dot-dashed line;
(K93, 2, 40, 3) green dotted line (almost overlapping the lower blue dashed line).
}  
\label{fig:fsg2} 
\end{figure}

From the results shown in Fig.\ref{fig:fsg1} one can easily
calculate \fsg as a function of $\phihalo$ and \fsgcs for any of the
combinations of IGCMF, cluster structural  
parameters and IMF explored.  For example,
Fig.\,\ref{fig:fsg2}, shows the dependence of \fsg on $\phihalo$
for a power-law IGCMF with
$\alpha=1.8$, for various values of $M_{low}$, $R$, $R_h$ and for
either a K93 or a K01 IMF (and, as discussed above, we have adopted
$\fsgcse=0.5$). 
If clusters were at the time of SG formation ($30\simlt t \simlt 100$
Myr) more compact than assumed here, the lines of \fsg vs $\phihalo$
would fall between those shown in Fig.\,\ref{fig:fsg2} and that corresponding to
the absolute upper limit easily obtained using  Eq.\ref{eq:fsgb} and the maximum
value of $\eta$ ($\eta_{max}\simeq 0.08$ for a K93 IMF and
$\eta_{max}\simeq 0.125$ for a K01 IMF).

Figs. \ref{fig:fsg1} and \ref{fig:fsg2} demonstrate that, for the
broad range of values of the IGCMF parameters considered, the
predicted fraction of SG stars in the halo is always small,
$\fsge<4-6\%$ for a K93 IMF and $\fsge<7-9\%$ for a K01 IMF . 
This result is consistent with the small fraction ($\sim$1.5-2.5\%) of
halo stars identified as SG stars by Carretta et al. (2010, based on
the stars' sodium abundance) and Martell and Grebel (2010, based on CN
and CH bandstrengths).  For comparison, 50\% of globular cluster stars
are identified as SG when the same classification criteria are adopted
(Kraft 1994, Carretta et al. 2010, Martell \& Grebel 2010).
We note that for these estimates of $\fsge$, using, for example, a
power-law IGCMF with $\alpha=1.8$ as a reference model, our
calculations (Fig.\,\ref{fig:fsg2}) with a K01 IMF (K93 IMF) imply
that a fraction ranging from  about 20\% to about 40\% (30\% to about
60\%) of the Galactic stellar halo must be composed of stars
originally formed in globular clusters.  Although this  
estimate is very uncertain and both more calculations and further
observational studies of SG stars in the halo will be required to
refine it, it is interesting to note how the fractions of SG stars in
globular clusters and in the Galactic stellar halo connect and
constrain both models for the formation and dynamical evolution of
multiple stellar generations in globular clusters and the formation
history of the Galactic halo.

{\bf Acknowledgments.}
EV and SM were supported in part by NASA grants NNX07AG95G, NNX08AH15G
and NNX10AD86G and by NSF grant AST-0708299. AD acknowledges financial
support from italian MIUR through grant PRIN 2007  
(prot. 2007JJC53X). FD was  supported by PRIN MIUR 2007 'Multiple
Stellar Populations in Globular Clusters: Census, Characterization
and Origin' (prot. n. 20075TP5K9).

\end{document}